\documentclass[aps,prl,preprint,groupedaddress,showpacs]{revtex4}
\begin{document}

\title{A phase transition model for metric fluctuations in vacuum}

\author{M. Mehrafarin}
\email{mehrafar@cic.aut.ac.ir}
\affiliation{Department of Physics, Amirkabir University of Technology, Tehran 15914,
Iran\\ Center for Theoretical Physics and Mathematics,
AEOI, Tehran 14399, Iran}

\begin{abstract}
Regarding metric fluctuations as 
generating {\it roughness} on the fabric of the otherwise smooth vacuum, it is shown that in its simplest form, the effect can be described by the scalar $\phi^4$ model. The model exhibits  a second order phase transition between a smooth
(low-temperature) phase and a rough (high-temperature) one, corroborating the absence of metric fluctuations at low energies. In the rough phase near the critical point, vacuum is characterized by a power-law
behavior for the fluctuating field with critical exponent $\beta \approx 0.33$.
\end{abstract}

\pacs{95.30.Sf, 04.90.+e, 04.60.-m}

\maketitle

By inducing fluctuations in the metric, quantum effects are generally expected
to play a principle role in the structure of spacetime at the Plank scale.
Apart from any residual effects they may have on larger scales, these
fluctuations become an inseparable part of the spacetime itself when
considering its structure down to such fine scales, i.e.,
at temperatures near as high as the Plank temperature, $T_P \sim 10^{19}\ GeV$.
Close to $T_P$, one may visualize the effect as giving rise to small scale
curvature ({\it roughness}) on the fabric of the otherwise smooth spacetime.
Motivated by this picture, we show that in vacuum such metric
fluctuations, in simplest form, can be modelled by the scalar $\phi^4$ theory.
The model exhibits a phase
transition at the renormalized Plank temperature $T_P(\infty)>T_P$, the
transition being of second order between
a smooth phase ($T<T_P(\infty)$) and a rough one ($T>T_P(\infty)$), thus corroborating the absence of metric fluctuations at low energies. In the
rough phase near
$T_P(\infty)$, vacuum is characterized by a power-law behavior for the
fluctuations with critical
exponent $\beta \approx 0.33$.

In the following, we shall appropriately phrase the vacuum in
presence (absence) of metric fluctuations as the real (ideal) vacuum.

In accordance with general relativity, we take the ideal vacuum to be a de
Sitter spacetime with the line element
\begin{equation}
ds^2=dt^2-e^{2Ht} (dr^2+r^2 d\theta^2+r^2 \sin^2\theta \ d\varphi^2), \ \ \
H=\sqrt {\frac {8 \pi G \rho}{3}}
\end{equation}
where $\rho$ is the energy density of the vacuum. Neglecting fluctuations, it
is naively expected that $\rho \sim m_P^4$, $m_P$ being the Plank mass which is
of the order of $10^{19}\ GeV$.

Our model Hamiltonian will be defined by the change in vacuum energy due to
metric fluctuations, which we now proceed to consider. Since
we are not interested in the dynamics of fluctuations we take t=const., thus
concentrating on the spatial cross section $dt=0$ of the vacuum spacetime near the Plank time.
Consequently, hereafter, by vacuum this spatial cross section is meant. Noting
that the ideal vacuum is Euclidean, we denote its cartesian coordinates by
$x^i \ (i=1,2,3)$. Metric fluctuations are to be regarded so as to
generate roughness on the fabric of this ideal smooth vacuum. One way
to model these fluctuations would, then, be to view their effect extrinsically and
represent the real vacuum by a hypersurface in a higher dimensional Eulidean space. In its simplest form, the effect of fluctuations
may described by a scalar field $\phi(x^i)$ so that the real vacuum can be
represented by the hypersurface $x^4=\phi(x^i)$ in a four
dimensional Euclidean space with cartesian coordinates $(x^i,x^4)$; the
hyperplane $x^4=0$
(corresponding to $\phi=0$) representing the ideal vacuum, of course. The local unit normal to the hypersurface
$x^4=\phi(x^i)$ will then be given by
\begin{equation}
[1+(\nabla \phi)^2]^{-1/2} (-\nabla \phi, 1)
\end{equation}
By projecting an element $d\upsilon$ of this hypersurface onto the
hyperplane $x^4=0$, it is readily shown that
\begin{equation}
d\upsilon=[1+(\nabla \phi)^2]^{1/2} \ d^3x  \label{one}
\end{equation}
This is the volume element of the real vacuum. In absence of fluctuations
($\phi=0$), (\ref{one}) reduces to the volume element of the ideal flat
vacuum, as
it should. Thus, assuming that real vacuum is not too rough ($|\nabla \phi|
\ll 1$), the contribution of (\ref{one}) to the excess vacuum energy due to
fluctuations will be
\begin{equation}
\int \frac {1}{2} \ \rho (\nabla \phi)^2 \ d^3x
\end{equation}

The integrand, which arises as a result of volume
expansion due to the metric fluctuations, constitutes the "kinetic term" of our
model Hamiltonian. The complete Hamiltonian
has to take into account the interaction
energy of the fluctuations, represented by the "potential term" $V(\phi)$, too.
Then, naturally assuming that these fluctuations are
scale invariant, their scaling behavior is to be studied via the
renormalization group techniques.
Indeed, it is this renormalizability requirement that fixes the form of
$V(\phi)$ unambiguously. Thus bearing in mind that our model is three
dimensional and that $V(\phi)$ should be even in $\phi$, we take
\begin{equation}
V(\phi)=A(T) \phi^2 + B(T) \phi^4, \ \ \ B(T)>0
\end{equation}
where $T$ is the temperature. Terms beyond $\phi^6$ have been omitted because
they represent irrelevant (non-renormalizable) interactions. Also, the $\phi^6$
term has been deliberately made irrelevant by taking $B(T)>0$ and, therefore,
ignored. This is because the
latter term pertains to tricritical behavior \cite{Blume, Lawrie} which does
not seem to be
related to our problem. Now, as is well known, the above potential
shows phase transition provided that
$A(T)$ vanishes at a certain critical temperature which is naturally to be
taken as $T_P$. This, being the temperature near which metric
fluctuations are naively expected to occur, is of course the trial critical
temperature; the true critical temparature will be given by the renormalization
of this trial value
due to the $\phi^4$ term. Noting that our model is only applicable in
the vicinity of the transition point, we can replace $B(T)$ by its value, $b$,
at $T_P$ and write
\begin{equation}
A(T)=\frac {1}{2} \ a(T-T_P)
\end{equation}
where the constant $a$ is taken to be negative because we want the
high (low) temperature phase to be the rough (smooth) phase.

Thus, the change in vacuum energy due to metric fluctuations that defines our
model Hamiltonian is given by
\begin{equation}
H[\phi]=\int d^3x\ \{\frac {1}{2} \ \rho (\nabla \phi)^2+
\frac {1}{2} \ a(T-T_P)  \phi^2 + b \phi^4\}
\end{equation}
where $a<0$ and $b>0$. This is the standard $\phi^4$ model which, as is well
known (see e.g. \cite{Zinn, Uzunov}), exhibits
a second order phase transition at the critical temperature $T_P(\infty)>T_P$
given by the zero of the renormalized value of $A(T)$.
The high temperature phase ($T>T_P(\infty)$) near the critical point is
characterized by the scale-invariant behavior
\begin{equation}
\phi \sim (T-T_P(\infty))^ \beta
\end{equation}
and is therefore rough, while in the low temperature
phase ($T<T_P(\infty)$) $\phi=0$ so that the vacuum is smooth. The
critical exponent $\beta$ is found from the standard
renormalization of the $\phi^4$ model \cite {Zinn, Uzunov} to have the approximate value
0.33.

Of course, in the above model, we have simply assumed that the fluctuating metric components somehow couple to produce a single fluctuating field $\phi(x)$; we have not considered the possible physical mechanisms (if any) behind the coupling. However, this assumtion, although it simplifies the resulting model considerably, is not fundamental to our approach.  What we are essentially pointing out here is that metric fluctuations can be described by phase transition models corroborating the fact that they show no gross effect at `low' energies.


\begin{thebibliography}{widest-label}
\bibitem{Blume}R. Blume, V. Emery and R. B. Grifiths, Phys. Rev. A {\bf
4}, 1071 (1971).
\bibitem{Lawrie}I. D. Lawrie and S. Sarbach, in {\it Phase transitions and
critical phenomena}, Vol. 9, edited by C. Domb and J. L. Lebowitz (Academic Press,
London, 1984).
\bibitem{Zinn}J. Zinn-Justin, {\it Quantum field theory and critical phenomena} (Clarendon
Press, Oxford, 1989).
\bibitem{Uzunov} D. I. Uzunov, {\it Introduction to the theory of critical
phenomena} (World Scientific, Singapore, 1993).
\end{thebibliography}
\end{document}